\documentclass[conference]{IEEEconf}

\usepackage{graphicx}
\usepackage{multirow}
\usepackage{titlesec}
\usepackage{balance}
\usepackage{stfloats}
\usepackage{booktabs}
\usepackage{amsmath,amssymb}
\usepackage{url}
\usepackage{xcolor}
\usepackage{hyperref}
\usepackage{caption}
\usepackage{listings}

\lstdefinelanguage{HCL}{
  keywords={resource, variable, output, module, data, locals, provider, terraform},
  keywordstyle=\bfseries,
  sensitive=true,
  comment=[l]{\#},
  commentstyle=\itshape,
  string=[b]",
  morestring=[b]',
}

\renewcommand\thesection{\arabic{section}}
\renewcommand\thesubsectiondis{\thesection.\arabic{subsection}}

\begin{document}

\title{\textbf{\Large IaC-Guard-V: A Verification Framework for LLM-Generated Infrastructure-as-Code Repairs\\}}

\author{Lokesh Chauhan$^{1}$\\
	\normalsize $^{1}$Independent Researcher, Dallas, USA\\
	\normalsize lokesh0186@gmail.com
}

\maketitle

\footnotetext{This work was conducted independently and does not relate to the author's position at Amazon. No Amazon resources, data, or confidential information were used.}

\begin{abstract}
Infrastructure-as-Code (IaC) misconfigurations are a leading cause of cloud security incidents, and Large Language Models (LLMs) are increasingly proposed as automated repair agents. While recent work has established detect-repair-verify workflows for general-purpose code, the trustworthiness of LLM-generated repairs for IaC remains uninvestigated. IaC presents unique verification challenges: security scanners may produce inconsistent results across tools and configurations, infrastructure semantics cannot be validated through unit tests, and provider-specific rules create a fragmented verification landscape. We present IaC-Guard-V, a verification-centered framework that evaluates AI-generated IaC repairs through four verification gates: syntactic validity, target-issue resolution, regression safety, and patch minimality. We construct a multi-technology benchmark of 70 real-world misconfigured Terraform and Kubernetes artifacts spanning 70 unique scanner rules across eight violation classes. The evaluation spans three repair strategies across three LLM families in 630 experimental runs. Our results reveal that while all models achieve 100\% syntactic validity, only 32--50\% of repairs pass full verification under single-shot prompting. Verification-guided iterative repair statistically significantly improves verified-fix rates across all models, achieving 68--92\% compared to 32--50\% for baselines. Notably, an open-source model with verification-guided repair outperforms the strongest commercial model without it at one-twelfth the cost per verified fix. Cross-technology analysis reveals that Kubernetes repairs achieve near-perfect rates for commercial models but require verification-guided iteration for open-source models, demonstrating that technology complexity and model capability jointly determine repair trustworthiness. We further find that structured prompting consistently degrades Terraform repair quality across all models, a counterintuitive result that challenges common assumptions about constrained LLM output. All artifacts are released for reproducibility.
\end{abstract}

\IEEEoverridecommandlockouts
\vspace{1.5ex}
\begin{keywords}
\itshape Infrastructure-as-Code; large language models; automated program repair; verification; software reliability; Terraform
\end{keywords}

\section{Introduction}

Infrastructure-as-Code (IaC) has become the standard mechanism for provisioning and managing cloud infrastructure, enabling teams to define servers, networks, and security policies through machine-readable configuration files in formats such as Terraform HCL and Kubernetes YAML. However, IaC misconfigurations remain a leading cause of cloud security incidents: the 2025 Cloud Security Report found that 65\% of organizations experienced a cloud-related security incident in the past year, with misconfigurations ranking among the top causes~\cite{checkpoint2025,unit42_2024}. Static analysis tools such as Checkov, KICS, and Terrascan can detect these misconfigurations, but remediation remains a manual, error-prone process that scales poorly across large infrastructure codebases.

Large Language Models (LLMs) have emerged as promising automated repair agents. In the broader domain of automated program repair (APR), LLM-based systems now fix hundreds of real-world bugs at costs below \$1 per fix~\cite{xia2024chatrepair}, and recent work has established detect--repair--verify (DRV) workflows that iteratively refine LLM-generated patches using test feedback~\cite{cheng2026drv}. Naturally, practitioners are increasingly exploring LLMs for IaC repair, prompting models to fix flagged misconfigurations and deploying the generated patches with minimal review.

This practice is dangerous. While LLM-generated IaC repairs may appear syntactically correct, their actual trustworthiness---whether they resolve the target issue without introducing new vulnerabilities---remains uninvestigated for IaC specifically.

\textbf{Motivating example.} Consider the Terraform configuration in Listing~\ref{lst:motivating}, which triggers Checkov rule CKV\_AWS\_8 (``Ensure that launch configuration EBS volumes are encrypted''). When prompted to fix this misconfiguration, Claude Sonnet~4.6 produces a syntactically valid repair that changes \texttt{associate\_public\_ip\_address} from \texttt{true} to \texttt{false}---a plausible-looking security improvement, but one that addresses the \textit{wrong attribute entirely}. The actual fix requires adding \texttt{encrypted = true} to the root block device configuration. Without automated verification, a practitioner reviewing this repair might accept it, believing the security issue was resolved. Our verification harness catches this failure at gate V2 (target-issue resolution), and the verification-guided repair loop feeds this feedback back to the model, which produces the correct fix on the second attempt.

\begin{figure}[!tbp]
\begin{lstlisting}[language=HCL,title={\footnotesize\textbf{(a) Original --- CKV\_AWS\_8: EBS volumes not encrypted}}]
resource "aws_launch_configuration" "lc" {
  image_id      = "ami-0c55b24b055c14ff6"
  instance_type = "t2.micro"
  associate_public_ip_address = true
  # Missing: root_block_device { encrypted = true }
}
\end{lstlisting}
\vspace{-6pt}
\begin{lstlisting}[language=HCL,title={\footnotesize\textbf{(b) LLM ``fix'' --- syntactically valid but WRONG}}]
resource "aws_launch_configuration" "lc" {
  image_id      = "ami-0c55b24b055c14ff6"
  instance_type = "t2.micro"
  associate_public_ip_address = false  # Changed wrong
  # attribute! EBS encryption still missing.
}
\end{lstlisting}
\vspace{-6pt}
\begin{lstlisting}[language=HCL,title={\footnotesize\textbf{(c) Correct fix --- found by verify-loop (attempt 2)}}]
resource "aws_launch_configuration" "lc" {
  image_id      = "ami-0c55b24b055c14ff6"
  instance_type = "t2.micro"
  associate_public_ip_address = true
  root_block_device {
    encrypted = true  # Actual fix for CKV_AWS_8
  }
}
\end{lstlisting}
\caption{Motivating example from benchmark item BM-0051. The LLM produces a syntactically valid repair (b) that changes the wrong attribute. The verification harness detects this at gate V2 (target-issue resolution), and the iterative repair loop produces the correct fix (c) on the second attempt.}
\label{lst:motivating}
\end{figure}

General-purpose DRV workflows~\cite{cheng2026drv} rely on unit tests and functional test suites for verification, an approach that does not transfer to IaC for three reasons. First, IaC artifacts lack executable test suites; verification must rely on static security scanners that may produce inconsistent results across tools and configurations. Second, IaC encodes provider-specific semantics (e.g., AWS resource attributes, Kubernetes pod security contexts) that vary across hundreds of resource types, making correct repair far more context-dependent than fixing a Python function. Third, IaC files frequently contain multiple resources, and a repair that fixes one misconfiguration may silently introduce regressions in others---a failure mode that syntactic validity alone cannot detect.

In this paper, we present \textbf{IaC-Guard-V}, a verification-centered framework that evaluates LLM-generated IaC repairs through four verification dimensions: three binary gates---(V1)~syntactic validity, (V2)~target-issue resolution, (V3)~regression safety---and one informational metric, (V4)~patch minimality. We define the \textit{verified-fix rate} (VFR) metric, which counts a repair as successful only when it passes all three binary gates, and construct a multi-technology benchmark of 70 real-world misconfigured artifacts (50 Terraform, 20 Kubernetes) spanning 70 unique scanner rules and eight violation classes.

We make the following contributions:

\begin{enumerate}
\item \textbf{Framework.} IaC-Guard-V, a verification-centered framework with four verification gates tailored to IaC-specific challenges, including a verification-guided iterative repair loop that feeds scanner feedback back to the LLM.

\item \textbf{Benchmark.} A multi-technology benchmark of 70 misconfigured IaC artifacts drawn from Checkov's test suite, with stratified selection across 70 unique rules and eight violation classes, accompanied by baseline scanner outputs for reproducible evaluation.

\item \textbf{Empirical study.} A comprehensive evaluation of 630 experimental runs across three LLM families (Claude Opus~4.6, Claude Sonnet~4.6, Llama~4 Maverick), three repair strategies (plain prompting, structured prompting, verification-guided iterative repair), and two IaC technologies (Terraform, Kubernetes).

\item \textbf{Key findings.} (a)~While all models achieve 100\% syntactic validity, only 32--50\% of repairs pass full verification under single-shot prompting. (b)~Verification-guided iterative repair achieves 68--92\% VFR, statistically significantly outperforming baselines. (c)~An open-source model with verification outperforms the strongest commercial model without it at a fraction of the cost per verified fix. (d)~Structured prompting consistently degrades Terraform repair quality---a counterintuitive finding. (e)~Kubernetes repairs achieve near-perfect rates for commercial models but require verification for open-source models, revealing that technology complexity and model capability jointly determine trustworthiness.
\end{enumerate}

All benchmark items, the verification harness, and experimental artifacts are released for reproducibility.\footnote{Replication package: \url{https://github.com/[redacted-for-review]}}

\section{Background and Related Work}

We organize related work into three areas: LLM-based automated program repair, IaC-specific repair and generation, and IaC benchmarks and defect taxonomies. Table~\ref{tab:positioning} positions our study against the most closely related work.

\subsection{LLM-Based Automated Program Repair}

Large language models have transformed automated program repair (APR). Xia et al.~\cite{xia2023practical} demonstrated that directly leveraging LLMs for program repair achieves competitive results, establishing a practical baseline for the field. Zhang et al.~\cite{zhang2024survey} provide the first systematic literature review of 189 LLM-based APR papers, categorizing utilization strategies across semantic bugs, security vulnerabilities, and other repair scenarios. Yang et al.~\cite{yang2025survey} propose a unified taxonomy grouping 62 systems into four paradigms: fine-tuning, prompting, procedural pipelines, and agentic frameworks.

Among individual systems, ChatRepair~\cite{xia2024chatrepair} pioneered conversation-driven repair, interleaving patch generation with test feedback to fix 162 of 337 Defects4J bugs at \$0.42 each. AlphaRepair~\cite{xia2022alpharepair} demonstrated that zero-shot LLM application can outperform supervised approaches. RepairLLaMA~\cite{silva2025repairllama} demonstrated that parameter-efficient fine-tuning of open-source models can match or exceed commercial models, fixing 144 Defects4J v2 bugs. RepairAgent~\cite{bouzenia2024repairagent} introduced autonomous agent-based repair using a finite state machine, repairing 164 bugs including 39 not fixed by prior techniques. FLAMES~\cite{lecong2026flames} uses semantic-guided search to reduce memory consumption by 83\% while maintaining repair effectiveness.

Most relevant to our work, Cheng~\cite{cheng2026drv} introduced the Detect--Repair--Verify (DRV) workflow for securing LLM-generated code, constructing a benchmark of runnable web applications and comparing single-pass and bounded iterative DRV variants. The iterative refinement paradigm has broader roots: Chen et al.~\cite{chen2024selfdebug} showed that LLMs can self-debug by explaining and correcting their own code, and Madaan et al.~\cite{madaan2023selfrefine} demonstrated that iterative self-feedback improves LLM outputs across diverse tasks. Tang et al.~\cite{tang2024exploration} formalized the exploration-exploitation tradeoff inherent in iterative code repair. While DRV establishes the iterative verification paradigm we build upon, it targets general-purpose code (PHP, JavaScript, Python) and relies on unit tests and functional test suites for verification---an approach that does not transfer to IaC, where infrastructure semantics cannot be validated through unit tests, security scanners may disagree on the same artifact, and provider-specific rules create a fragmented verification landscape. IaC-Guard-V addresses these gaps by defining four IaC-specific verification gates and evaluating repair across multiple IaC technologies and scanner rules.

\subsection{IaC-Specific Repair and Generation}

Several recent works address IaC repair and generation directly. TerraFormer~\cite{jana2026terraformer} combines supervised fine-tuning with verifier-guided reinforcement learning (GRPO) for Terraform generation and mutation, demonstrating that a 14B fine-tuned model can outperform 50$\times$ larger models on correctness and security compliance. However, TerraFormer focuses on generation (NL-to-IaC) rather than repair of existing misconfigurations, and its verification is limited to policy compliance scoring without regression analysis.

InfraFix~\cite{saavedra2025infrafix} proposes the first technology-agnostic APR framework for IaC, using SMT-based repair guided by inferred system states across Ansible, Puppet, Chef, and Terraform. While InfraFix achieves 95.7\% success across 254,288 repair scenarios, it uses constraint solving rather than LLMs and does not evaluate the trustworthiness of LLM-generated repairs.

For Kubernetes specifically, GenKubeSec~\cite{malul2024genkubesec} fine-tunes a local LLM for misconfiguration detection and remediation across 169 misconfiguration types, achieving 0.990 precision on $\sim$277,000 configurations. LLMSecConfig~\cite{ye2025llmsecconfig} combines Checkov with LLMs and RAG for container misconfiguration repair, reporting 94.3\% success with Mistral Large 2 on 1,000 real-world configurations. Both focus on single-technology repair without cross-technology comparison or multi-dimensional verification.

IaCGen~\cite{sun2025iacgen} introduces a deployability-centric benchmark (153 AWS scenarios) and an iterative feedback framework achieving 54.6--91.6\% deployment success within 10 iterations. Palavalli and Santolucito~\cite{palavalli2024feedback} study feedback loops for CloudFormation generation, finding that effectiveness decreases exponentially and plateaus after approximately 5 iterations---a finding our convergence analysis corroborates for repair (Section~VI-E).

\subsection{IaC Benchmarks and Defect Taxonomies}

The IaC research community has established several benchmarks and taxonomies. Rahman et al.~\cite{rahman2019sevensins} identified seven categories of security smells in IaC scripts through analysis of 1,726 Puppet scripts, establishing the foundational taxonomy that tools like Checkov build upon. Sharma et al.~\cite{sharma2016smells} cataloged 24 configuration smells across 4,621 Puppet repositories, while Steffens et al.~\cite{steffens2018smells} extended this work to Chef with 17 technology-agnostic smells. Saavedra and Ferreira~\cite{saavedra2022glitch} proposed GLITCH, a polyglot framework for cross-technology security smell detection across Ansible, Chef, Puppet, and Terraform. Dalla Palma et al.~\cite{dallapalma2022defect} developed ML-based defect prediction for IaC using product and process metrics. Verdet et al.~\cite{verdet2025checkov} conducted the first large-scale empirical study of Terraform security practices using Checkov and tfsec across 812 projects, finding significant variation in security policy adoption across cloud providers. Hummer et al.~\cite{hummer2013idempotence} proposed model-based testing for IaC idempotence.

For benchmarking LLM capabilities, IaC-Eval~\cite{kon2024iaceval} evaluates Terraform code generation with formal verification, showing that GPT-4 achieves only 19.36\% pass@1 accuracy---far below its 86.6\% on Python benchmarks---highlighting the difficulty gap between general code and IaC. Rahman et al.~\cite{rahman2023k8s} conducted a large-scale empirical study of security misconfigurations in open-source Kubernetes manifests, providing foundational data on defect prevalence.

\textbf{Gap.} No prior work combines LLM-based repair, multi-dimensional verification (syntax, resolution, regression, minimality), and cross-technology IaC evaluation in a single framework. IaC-Guard-V fills this gap.

\begin{table*}[!tbp]
\centering
\caption{Positioning of This Study Against Related Work}
\label{tab:positioning}
\begin{tabular}{|l|l|c|c|c|c|c|}
\hline
\textbf{Study} & \textbf{Focus} & \textbf{IaC} & \textbf{LLM} & \textbf{Verification} & \textbf{Repair} & \textbf{Benchmark} \\
\hline
DRV~\cite{cheng2026drv} & General code DRV & No & Yes & Unit tests & Yes & Yes (web apps) \\
ChatRepair~\cite{xia2024chatrepair} & Conversational APR & No & Yes & Unit tests & Yes & Defects4J \\
TerraFormer~\cite{jana2026terraformer} & IaC generation + RL & Yes & Yes & Policy score & No (gen) & Yes (152K) \\
InfraFix~\cite{saavedra2025infrafix} & IaC repair (SMT) & Yes & No & Partial & Yes & Yes (254K) \\
GenKubeSec~\cite{malul2024genkubesec} & K8s misconfig repair & Yes & Yes & Detection only & Yes & Yes (277K) \\
LLMSecConfig~\cite{ye2025llmsecconfig} & Container repair & Yes & Yes & Checkov & Yes & Yes (1K) \\
IaC-Eval~\cite{kon2024iaceval} & IaC generation bench & Yes & Yes & Formal & No & Yes (458) \\
IaCGen~\cite{sun2025iacgen} & IaC gen + feedback & Yes & Yes & Deploy test & No (gen) & Yes (153) \\
\hline
\textbf{This study} & \textbf{IaC repair verif.} & \textbf{Yes} & \textbf{Yes} & \textbf{4 gates} & \textbf{Yes} & \textbf{Yes (70)} \\
\hline
\end{tabular}
\end{table*}

\section{IaC-Guard-V Framework}

\begin{figure*}[!tbp]
\centering
\includegraphics[width=0.85\textwidth]{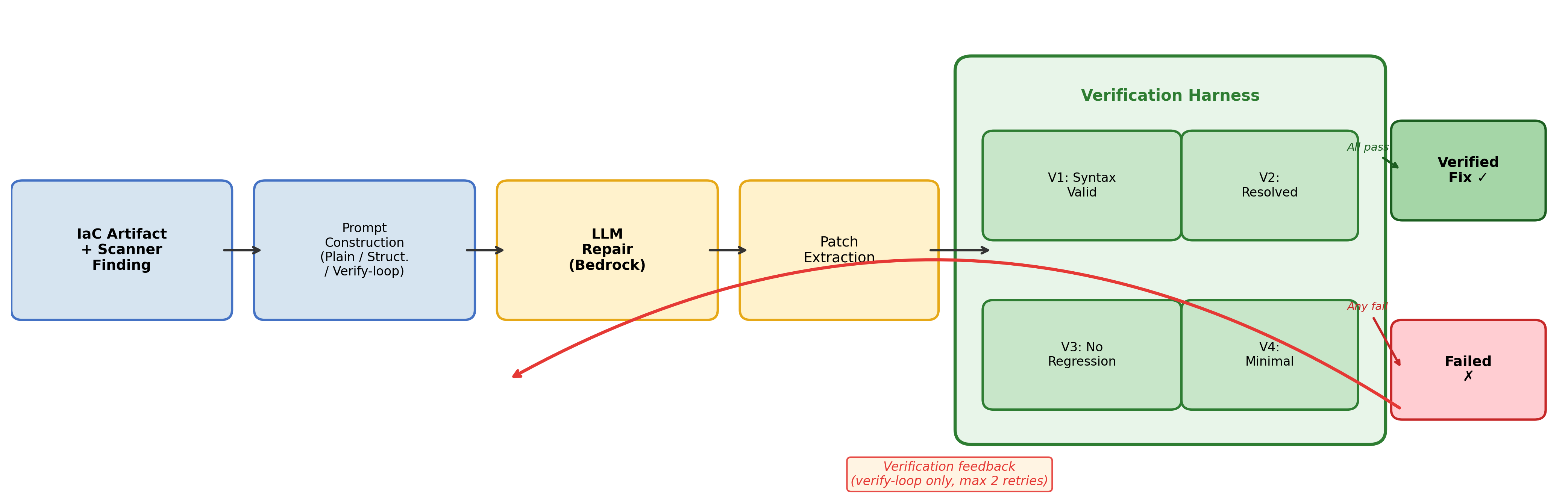}
\caption{IaC-Guard-V pipeline. An IaC artifact with a scanner finding is repaired by an LLM using one of three strategies. The verification harness evaluates the repair across four gates. In verify-loop mode, failed verification results are fed back to the LLM for up to two retries.}
\label{fig:pipeline}
\end{figure*}

This section presents IaC-Guard-V, a verification-centered framework for evaluating LLM-generated IaC repairs. Figure~\ref{fig:pipeline} illustrates the end-to-end pipeline.

\subsection{Verification Gates}

Given an original IaC artifact $A$, a repaired artifact $A' = f(A, v)$ produced by an LLM in response to a scanner finding $v$, we define a repair as a \textit{verified fix} if and only if it passes all three binary verification gates:

\begin{equation}
\mathit{VF}(A, A', v) = V_1(A') \wedge V_2(A', v) \wedge V_3(A, A')
\end{equation}

\noindent Patch minimality ($V_4$) is measured but does not gate the verdict, as there is no universally accepted threshold for acceptable patch size.

\textbf{V1: Syntactic Validity.} The repaired artifact must be parseable by the target technology's parser (HCL for Terraform, YAML for Kubernetes). We use Checkov's built-in parser as the syntax check: if Checkov can scan the file and produce structured output, the artifact is syntactically valid.

\textbf{V2: Target-Issue Resolution.} The specific scanner finding $v$ that triggered the repair must no longer appear in the Checkov output for $A'$. We re-run Checkov on the repaired artifact and verify that the target rule ID is absent from the failed-checks list. In multi-resource files, we check the rule at the file level; a limitation of this approach is discussed in Section~VI-E.

\textbf{V3: Regression Safety.} The repair must not introduce new scanner findings. We compare the set of failed Checkov rule IDs on $A'$ against the baseline set on $A$. Any rule ID that appears in the repaired output but not in the baseline constitutes a regression, regardless of which resource triggers it. A verified fix requires zero regressions.

\textbf{V4: Patch Minimality.} We measure the size of the repair as the number of lines added, removed, or modified (computed via unified diff). Unlike V1--V3, minimality is not a binary gate but an informational metric: excessively large patches suggest the LLM rewrote unrelated sections rather than making a targeted fix. We report minimality statistics to characterize repair precision.

\subsection{Verified-Fix Rate}

The primary evaluation metric is the \textit{verified-fix rate} (VFR):

\begin{equation}
\mathit{VFR} = \frac{|\{i : V_1(A'_i) \wedge V_2(A'_i, v_i) \wedge V_3(A_i, A'_i)\}|}{N}
\end{equation}

\noindent where $N$ is the total number of benchmark items. VFR is strictly more conservative than metrics that count syntactically valid or plausible-looking repairs, as it requires simultaneous satisfaction of syntax, resolution, and regression gates.

\subsection{Repair Strategies}

We evaluate three repair strategies of increasing sophistication:

\textbf{Plain prompting.} The LLM receives the misconfigured artifact and a generic instruction to fix the security misconfiguration. No information about the specific rule, affected resource, or expected fix is provided. The model returns raw IaC code.

\textbf{Structured prompting.} The LLM receives the artifact along with the Checkov rule ID, rule description, violation class, affected resource name, and affected line range. The model is instructed to return a JSON object containing the fixed artifact, a patch summary, and lists of changed resources and attributes. This strategy tests whether providing structured context and constraining output format improves repair quality.

\textbf{Verification-guided iterative repair.} The first attempt uses the structured prompt because it provides the LLM with maximum context about the target rule, even though single-shot structured prompting underperforms plain prompting (Section~VI-A). If verification fails, the harness constructs a retry prompt containing the original artifact, the failed attempt, and detailed verification feedback (which gates passed, which failed, and the Checkov output on the failed repair). The LLM receives this feedback and produces a revised fix. We allow up to two retries (three total attempts), chosen based on preliminary experiments showing diminishing returns beyond the second retry (confirmed in Section~VI-E). This strategy tests whether verification feedback enables the LLM to self-correct.

\textbf{Patch extraction.} For plain prompting, we extract the response text, stripping any markdown code fences. For structured and verification-guided prompting, we parse the JSON response and extract the \texttt{fixed\_artifact} field; if JSON parsing fails, we fall back to regex extraction of the JSON object.

\subsection{Implementation}

The IaC-Guard-V framework is implemented in Python (approximately 1,200 lines of code) and consists of four modules. The \textit{prompt constructor} generates method-specific prompts by combining the original artifact with Checkov metadata (rule ID, description, affected resource, line range). The \textit{Bedrock caller} handles API communication with AWS Bedrock, supporting both Anthropic (Messages API) and Meta (Llama) model families with automatic retry on transient errors. The \textit{verification harness} orchestrates the four-gate evaluation pipeline: it invokes Checkov as a subprocess, parses the JSON output to extract failed rules, computes the diff between original and repaired artifacts, and renders a per-gate verdict. The \textit{experiment runner} coordinates the full evaluation loop, managing the verify-loop retry logic and recording all intermediate results (per-attempt verification, token counts, latency) to JSON files for post-hoc analysis.

The verification harness executes Checkov in quiet mode with JSON output (\texttt{checkov -f <file> --output json --quiet --compact}), parsing the structured results to extract the list of failed rule IDs. For V2, we check whether the target rule ID appears in the failed-checks list of the repaired file. For V3, we compute the set difference between the repaired file's failed rules and the baseline's failed rules; any new rule constitutes a regression. For V4, we compute a unified diff and count added, removed, and modified lines, reporting the diff ratio (changed lines divided by original file length) as the minimality metric.

\section{Benchmark Construction}

We construct a multi-technology benchmark of 70 misconfigured IaC artifacts designed to evaluate repair across diverse violation types and technologies.

\subsection{Source and Selection}

We source benchmark items from Checkov's official test suite (v3.2.517), which contains IaC artifacts written by the tool's maintainers to validate specific scanner rules. These artifacts represent realistic misconfiguration patterns---each is a minimal, self-contained file that triggers one or more known Checkov rules---while being publicly available and free of proprietary content.

For Terraform, we scanned 287 AWS-specific test directories containing 1,081 individual rule violations across 369 unique Checkov rules. We selected 50 items using stratified sampling: one item per unique rule, with deliberate over-sampling of minority violation classes (e.g., public exposure at 12\% of the benchmark vs.\ 4\% of the corpus) to avoid inflating fix rates with easily repaired encryption misconfigurations, which dominate the corpus at 29\%. For Kubernetes, we scanned 98 test directories containing 2,400 violations across 215 unique rules and selected 20 items using the same one-per-rule strategy.

\subsection{Violation Taxonomy}

We classify benchmark items into eight violation classes based on the misconfiguration's security impact. Table~\ref{tab:benchmark} summarizes the benchmark composition.

\begin{table}[!tbp]
\centering
\caption{Benchmark Composition}
\label{tab:benchmark}
\begin{tabular}{|l|c|c|}
\hline
\textbf{Violation Class} & \textbf{Terraform} & \textbf{Kubernetes} \\
\hline
Missing encryption & 12 & 2 \\
Over-permissive access & 8 & 6 \\
Network hardening & 8 & 3 \\
Public exposure & 6 & --- \\
Weak observability & 6 & 1 \\
Insecure defaults & 5 & 3 \\
Missing runtime safety & --- & 4 \\
Other & 5 & 1 \\
\hline
\textbf{Total} & \textbf{50} & \textbf{20} \\
\hline
\multicolumn{3}{l}{\footnotesize Each item maps to a unique Checkov rule (70 total).} \\
\end{tabular}
\end{table}

\subsection{Baseline Establishment}

For each benchmark item, we run Checkov and record the complete set of failed rules as the baseline. This baseline serves two purposes: (1)~confirming that the target rule is present in the scanner output (verified for all 70 items), and (2)~providing the reference set for regression detection (V3). Baseline Checkov outputs, raw artifacts, and the selection manifest are included in the replication package.

\section{Experimental Setup}

\subsection{Models}

We evaluate three LLMs spanning two model families, accessed via the AWS Bedrock API: Claude Opus~4.6 and Claude Sonnet~4.6 (Anthropic, commercial, undisclosed parameter counts) and Llama~4 Maverick~17B (Meta, open-weight, 17B active parameters in a mixture-of-experts architecture). This selection provides two commercial models of different capability levels (Opus, Sonnet) and one open-source model (Maverick), enabling analysis of how model capability and openness affect repair quality.

\subsection{Configuration}

All models use temperature~0 for deterministic outputs, ensuring reproducibility without repeated runs. Maximum output length is set to 4,096 tokens. For verification-guided iterative repair, we allow up to two retries (three total attempts). All verification runs use Checkov v3.2.517. Prompts are provided in the replication package.

\subsection{Evaluation Protocol}

Each of the 70 benchmark items is evaluated with each of the three repair strategies on each of the three models, yielding $70 \times 3 \times 3 = 630$ experimental runs. For each run, we record: the raw model response, the extracted patch, all four verification dimension results, token counts (input and output), and wall-clock latency. For verification-guided runs, we additionally record per-attempt verification results to enable convergence analysis.

\subsection{Statistical Methods}

We report 95\% bootstrap confidence intervals (1,000 resamples with fixed seed) for all verified-fix rates. We test for overall method differences using Cochran's Q test and perform six pre-specified pairwise comparisons (verify-loop vs.\ plain and verify-loop vs.\ structured, for each of the three models on the Terraform benchmark) using McNemar's exact test with Bonferroni correction ($\alpha = 0.05/6 = 0.0083$). We report Cliff's delta as the effect size measure alongside each p-value. Kubernetes comparisons are reported descriptively, as commercial models achieve ceiling performance (100\%).

\section{Results}

We organize our results around five research questions. All experiments use temperature 0 for deterministic outputs; each benchmark item is evaluated once per model-method combination, yielding 630 total runs (450 Terraform, 180 Kubernetes).

\subsection{RQ1: Effectiveness of Verification-Guided Repair}

\textit{Motivation.} Our central claim is that verification-guided iterative repair produces more trustworthy fixes than single-shot prompting. We evaluate this on the Terraform benchmark (50 items, 3 models, 3 methods).

\textit{Approach.} For each benchmark item, we run all three repair strategies and measure the verified-fix rate (VFR): the proportion of repairs that pass all three binary verification gates. We report 95\% bootstrap confidence intervals (1,000 resamples) and test significance using McNemar's test with Bonferroni correction ($\alpha = 0.05/6 = 0.0083$ for six pre-specified comparisons).

\begin{figure}[!tbp]
\centering
\includegraphics[width=\columnwidth]{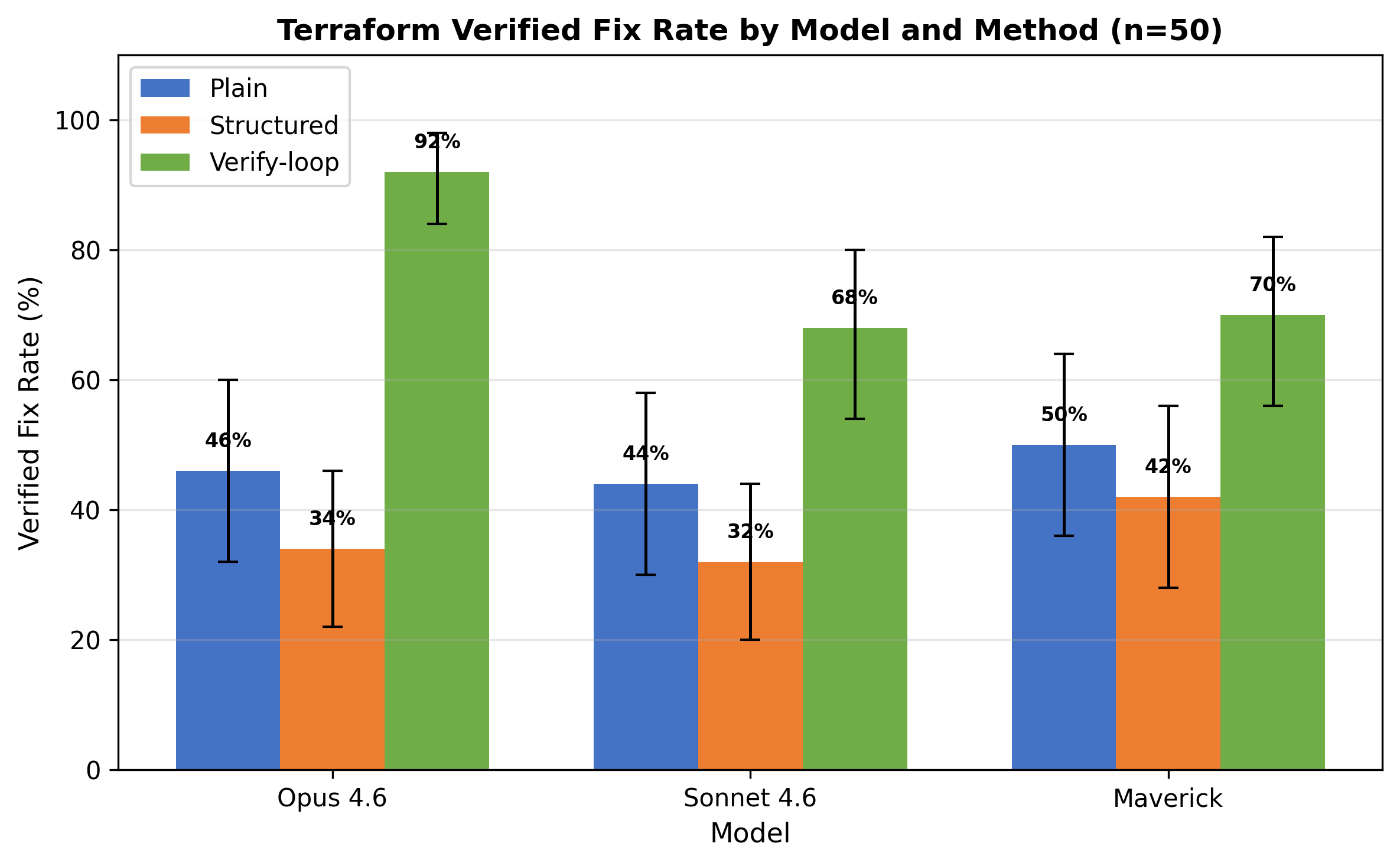}
\caption{Verified-fix rate by model and repair strategy (Terraform, n=50). Verify-loop substantially outperforms both baselines for all models. Error bars show 95\% bootstrap confidence intervals.}
\label{fig:hero_vfr}
\end{figure}

\textit{Results.} Table~\ref{tab:terraform_results} and Figure~\ref{fig:hero_vfr} present the Terraform results. Three findings emerge.

First, all models achieve 100\% syntactic validity across all methods, confirming that modern LLMs reliably generate parseable HCL. However, syntactic validity alone is a poor proxy for correctness: only 34--52\% of syntactically valid repairs resolve the target issue under single-shot prompting.

Second, verification-guided iterative repair (verify-loop) substantially outperforms both baselines for every model. Opus achieves 92\% VFR (CI: 84--98\%), Maverick 70\% (CI: 56--82\%), and Sonnet 68\% (CI: 54--80\%), compared to 32--50\% for plain and structured prompting. The confidence intervals for verify-loop do not overlap with those of structured prompting for any model.

Third, structured prompting consistently underperforms plain prompting (32--42\% vs.\ 44--50\%), a counterintuitive finding we discuss in Section~VII. Structured prompting also introduces regressions in 4--10\% of cases, while plain prompting on Opus and Sonnet introduces none.

\begin{table}[!tbp]
\centering
\caption{Terraform Repair Results (n=50 per cell)}
\label{tab:terraform_results}
\begin{tabular}{|l|l|c|c|c|c|c|}
\hline
\textbf{Model} & \textbf{Method} & \textbf{V1} & \textbf{V2} & \textbf{V3} & \textbf{VFR} & \textbf{CI} \\
\hline
\multirow{3}{*}{Opus} & Plain & 100 & 46 & 100 & \textbf{46} & 32--60 \\
 & Struct. & 100 & 34 & 96 & 34 & 22--46 \\
 & V-loop & 100 & 92 & 98 & \textbf{92} & 84--98 \\
\hline
\multirow{3}{*}{Sonnet} & Plain & 100 & 44 & 100 & \textbf{44} & 30--58 \\
 & Struct. & 100 & 34 & 90 & 32 & 20--44 \\
 & V-loop & 100 & 76 & 92 & \textbf{68} & 54--80 \\
\hline
\multirow{3}{*}{Mav.} & Plain & 100 & 52 & 88 & \textbf{50} & 36--64 \\
 & Struct. & 100 & 42 & 100 & 42 & 28--56 \\
 & V-loop & 100 & 74 & 94 & \textbf{70} & 56--82 \\
\hline
\end{tabular}

{\footnotesize V1=Syntax, V2=Resolved, V3=No Regression (\%). CI=95\% bootstrap. V4 (minimality) reported separately.}
\end{table}

Statistical significance confirms these findings. Cochran's Q test rejects the null hypothesis of equal method performance for all three models ($p < 0.001$). In pairwise McNemar's tests with Bonferroni correction ($\alpha = 0.0083$), verify-loop significantly outperforms structured prompting for all three models: Opus ($p < 0.001$, Cliff's $\delta = 0.58$, large), Sonnet ($p < 0.001$, $\delta = 0.36$, medium), and Maverick ($p < 0.001$, $\delta = 0.28$, small). Verify-loop versus plain is significant for Opus ($p < 0.001$, $\delta = 0.46$, medium) and approaches significance for Sonnet ($p = 0.008$, $\delta = 0.24$) and Maverick ($p = 0.006$, $\delta = 0.20$); all three are significant under Holm-Bonferroni correction.

\subsection{RQ2: Variation by Violation Class}

\textit{Motivation.} Not all misconfigurations are equally difficult to repair. We analyze whether verified-fix rates vary by violation class.

\textit{Approach.} We aggregate VFR across all three models for each violation class and compare plain versus verify-loop to identify where verification feedback provides the greatest benefit.

\textit{Results.} Table~\ref{tab:violation_class} presents verify-loop VFR by violation class, aggregated across all three models. Fix rates range from 47\% (insecure defaults) to 97\% (missing encryption). Missing-encryption violations are the easiest to repair because they typically require adding a single attribute (e.g., \texttt{encrypted = true}). Insecure-defaults and ``other'' violations are hardest because they involve multi-resource files or provider-specific semantics that LLMs struggle to resolve. Public-exposure violations show the largest improvement from verify-loop over plain (94\% vs.\ 17\%), suggesting that verification feedback is most valuable for violations where the initial fix attempt is likely to be incomplete.

\begin{table}[!tbp]
\centering
\caption{Verify-Loop VFR by Violation Class (Terraform)}
\label{tab:violation_class}
\begin{tabular}{|l|c|c|c|}
\hline
\textbf{Violation Class} & \textbf{Plain} & \textbf{Verify-loop} & \textbf{$\Delta$} \\
\hline
Missing encryption & 83 & \textbf{97} & +14 \\
Public exposure & 17 & \textbf{94} & +78 \\
Weak observability & 56 & \textbf{83} & +28 \\
Over-permissive access & 46 & \textbf{71} & +25 \\
Network hardening & 33 & \textbf{67} & +33 \\
Other & 20 & \textbf{53} & +33 \\
Insecure defaults & 33 & \textbf{47} & +13 \\
\hline
\multicolumn{4}{l}{\footnotesize All values are percentages, aggregated across 3 models.} \\
\end{tabular}
\end{table}

Table~\ref{tab:permodel_class} disaggregates verify-loop VFR by model and violation class, revealing substantial model-dependent variation. Opus achieves 100\% on missing encryption and insecure defaults, while Sonnet and Maverick achieve only 20\% on insecure defaults---an 80 percentage point gap within the same violation class. The ``other'' category shows the most extreme divergence: Opus and Maverick achieve 80\% while Sonnet achieves 0\%, suggesting that Sonnet's repair strategy fails on misconfiguration types that fall outside well-defined categories. These per-model differences underscore the importance of evaluating across multiple models rather than drawing conclusions from a single LLM.

\begin{table}[!tbp]
\centering
\caption{Verify-Loop VFR by Model and Violation Class (Terraform, \%)}
\label{tab:permodel_class}
\begin{tabular}{|l|c|c|c|}
\hline
\textbf{Violation Class} & \textbf{Opus} & \textbf{Sonnet} & \textbf{Maverick} \\
\hline
Missing encryption (12) & \textbf{100} & \textbf{100} & 92 \\
Public exposure (6) & \textbf{100} & 83 & \textbf{100} \\
Weak observability (6) & 83 & 83 & 83 \\
Over-permissive access (8) & \textbf{88} & 75 & 50 \\
Network hardening (8) & \textbf{88} & 63 & 50 \\
Other (5) & 80 & 0 & 80 \\
Insecure defaults (5) & \textbf{100} & 20 & 20 \\
\hline
\textbf{Overall (50)} & \textbf{92} & 68 & 70 \\
\hline
\multicolumn{4}{l}{\footnotesize Numbers in parentheses indicate items per class.} \\
\end{tabular}
\end{table}

\subsection{RQ3: Cross-Technology Generalization}

\textit{Motivation.} To evaluate whether our findings generalize beyond Terraform, we extend the benchmark with 20 Kubernetes manifest items spanning 20 unique Checkov rules across six violation classes.

\textit{Results.} Table~\ref{tab:k8s_results} presents the Kubernetes results. Both commercial models (Sonnet and Opus) achieve 100\% VFR across all three methods, while Maverick achieves only 45\% with plain prompting before verification-guided iteration rescues it to 100\%.

\begin{table}[!tbp]
\centering
\caption{Kubernetes Repair Results (n=20 per cell)}
\label{tab:k8s_results}
\begin{tabular}{|l|l|c|c|c|c|}
\hline
\textbf{Model} & \textbf{Method} & \textbf{V1} & \textbf{V2} & \textbf{V3} & \textbf{VFR} \\
\hline
\multirow{3}{*}{Opus} & Plain & 100 & 100 & 100 & \textbf{100} \\
 & Struct. & 100 & 100 & 100 & \textbf{100} \\
 & V-loop & 100 & 100 & 100 & \textbf{100} \\
\hline
\multirow{3}{*}{Sonnet} & Plain & 100 & 100 & 100 & \textbf{100} \\
 & Struct. & 100 & 100 & 100 & \textbf{100} \\
 & V-loop & 100 & 100 & 100 & \textbf{100} \\
\hline
\multirow{3}{*}{Mav.} & Plain & 100 & 70 & 55 & \textbf{45} \\
 & Struct. & 95 & 95 & 95 & \textbf{95} \\
 & V-loop & 100 & 100 & 100 & \textbf{100} \\
\hline
\end{tabular}

{\footnotesize All values are percentages. VFR requires passing all gates simultaneously.}
\end{table}

This model-dependent disparity reveals that technology simplicity alone does not guarantee repair success; model capability remains a factor. Notably, Maverick is the only model to produce syntactically invalid output (V1=95\% on Kubernetes structured), suggesting that the JSON schema constraint occasionally disrupts YAML generation for open-source models. Structured prompting dramatically improves Maverick's Kubernetes performance (45\% $\rightarrow$ 95\%), the opposite of its effect on Terraform where it consistently degrades quality. We attribute the overall technology gap to three structural differences: (1) Kubernetes manifests are declarative YAML with flat, well-documented schemas, whereas Terraform HCL encodes provider-specific semantics across hundreds of resource types; (2) Terraform misconfigurations frequently involve cross-resource dependencies absent in typical Kubernetes manifests; (3) Terraform test files often contain multiple intentionally misconfigured resources, requiring selective repair without disturbing others.

\subsection{RQ4: Cost-Effectiveness}

\textit{Motivation.} Verification-guided repair requires multiple LLM invocations, increasing cost and latency. We quantify this tradeoff using actual token counts and AWS Bedrock pricing (April 2026).

\textit{Results.} Table~\ref{tab:cost} presents cost-effectiveness metrics. Maverick with verify-loop achieves 78.6\% VFR at \$0.002 per verified fix and 5.0s average latency. Opus without verification achieves 61.4\% at \$0.029 per verified fix---nearly 12$\times$ more expensive per successful repair. This demonstrates that an open-source model with verification produces more verified fixes per dollar than any commercial model configuration. Sonnet verify-loop achieves comparable quality to Maverick (77.1\% vs.\ 78.6\%) but at 18$\times$ higher cost per fix (\$0.044 vs.\ \$0.002) and 3.5$\times$ higher latency (17.6s vs.\ 5.0s).

\begin{table}[!tbp]
\centering
\caption{Cost-Effectiveness (All 70 Items)}
\label{tab:cost}
\begin{tabular}{|l|l|c|c|c|c|}
\hline
\textbf{Model} & \textbf{Meth.} & \textbf{VFR} & \textbf{Tok.} & \textbf{Lat.} & \textbf{\$/fix} \\
\hline
Mav. & Plain & 48.6 & 1,027 & 1.7s & .001 \\
Mav. & Verify & \textbf{78.6} & 3,163 & 5.0s & .002 \\
Sonnet & Plain & 60.0 & 1,217 & 5.1s & .018 \\
Sonnet & Verify & 77.1 & 4,188 & 17.6s & .044 \\
Opus & Plain & 61.4 & 1,214 & 6.4s & .029 \\
Opus & Verify & \textbf{94.3} & 3,320 & 15.2s & .047 \\
\hline
\end{tabular}

{\footnotesize Tok.=avg tokens. Lat.=avg latency. \$/fix=USD per verified fix. VFR across 50 TF + 20 K8s items. Structured omitted (underperforms plain).}
\end{table}

\subsection{RQ4b: Patch Minimality (V4)}

\textit{Motivation.} A correct repair that rewrites the entire file is less desirable than a targeted, minimal patch. We analyze patch minimality across methods and models using the diff ratio (changed lines / original lines).

\textit{Results.} Table~\ref{tab:minimality} presents median diff ratios for Terraform repairs. Three patterns emerge. First, structured prompting produces the most minimal patches (median 0.067--0.086) across all models, likely because the JSON schema constraint forces the model to articulate specific changes rather than regenerating the entire file. Second, Maverick produces substantially larger patches than the Claude models under plain prompting (median 0.478 vs.\ 0.162--0.227), suggesting that the open-source model is more prone to rewriting unrelated sections. Third, verify-loop patches are slightly larger than plain patches for Claude models (0.108--0.152 vs.\ 0.162--0.227) because retry attempts sometimes modify additional resources to satisfy verification. However, Maverick's verify-loop patches (0.144) are substantially \textit{smaller} than its plain patches (0.478), indicating that verification feedback helps the open-source model focus its repairs.

\begin{table}[!tbp]
\centering
\caption{Patch Minimality: Median Diff Ratio (Terraform)}
\label{tab:minimality}
\begin{tabular}{|l|c|c|c|}
\hline
\textbf{Model} & \textbf{Plain} & \textbf{Structured} & \textbf{Verify-loop} \\
\hline
Opus 4.6 & 0.227 & \textbf{0.067} & 0.152 \\
Sonnet 4.6 & 0.162 & \textbf{0.073} & 0.108 \\
Maverick & 0.478 & \textbf{0.086} & 0.144 \\
\hline
\multicolumn{4}{l}{\footnotesize Diff ratio = changed lines / original lines. Lower is more minimal.} \\
\end{tabular}
\end{table}

This reveals a tradeoff: structured prompting optimizes for minimality but sacrifices correctness (32--42\% VFR), while verify-loop optimizes for correctness (68--92\% VFR) at the cost of slightly larger patches. For practical deployment, correctness should take priority, but the minimality data suggests that combining structured output constraints with verification feedback could yield both minimal and correct repairs---a direction for future work.

\subsection{RQ5: Failure Analysis}

\textit{Motivation.} Understanding where and why repairs fail is essential for improving both models and verification harnesses.

\textit{Results.} We analyze failures across all Terraform verify-loop runs. Of the 50 benchmark items, 30 are fixed by all three models, 4 are fixed by none, and 16 show model-dependent outcomes. The 16 model-dependent items produce 35 individual model-item failures across the three models, which we categorize into three types:

\textbf{Failure categorization.} The author manually inspected all 35 verify-loop failures, examining the original artifact, the LLM's repair attempts, and the Checkov output for each attempt. We identify three categories: (1) two items (BM-0276, BM-0449) where the verification harness cannot isolate the target resource from pre-existing failures in multi-resource files---these are harness limitations, not model failures; (2) one item (BM-0040) involving a graph-based Checkov rule (CKV2\_AWS\_33) that checks cross-resource dependencies, which no model resolves; (3) the remaining failures are model limitations where the LLM fails to produce a correct fix despite retries.

\begin{figure}[!tbp]
\centering
\includegraphics[width=\columnwidth]{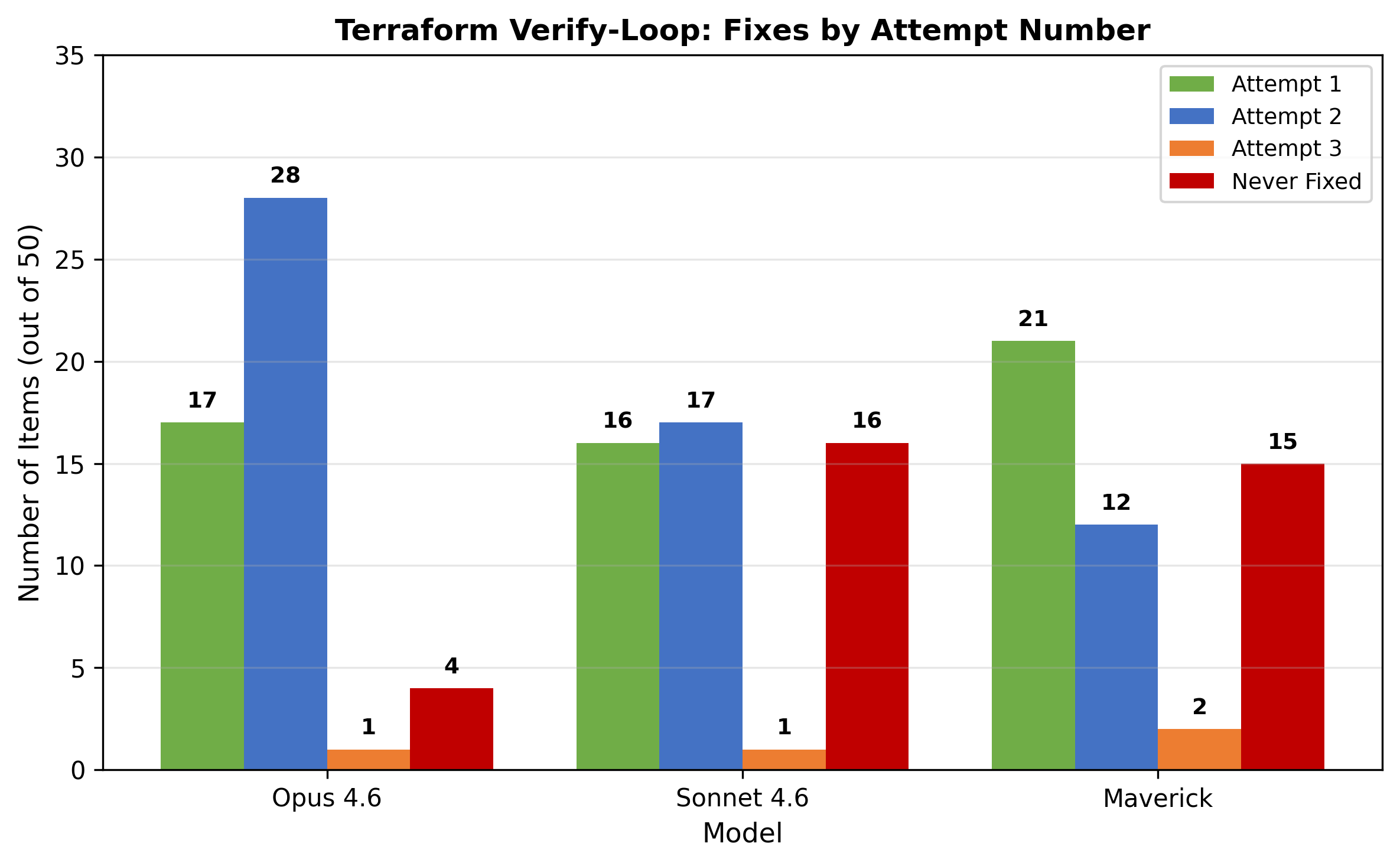}
\caption{Terraform verify-loop: distribution of fixes by attempt number. Most fixes occur in attempts 1--2; a third retry adds only 2 percentage points on average.}
\label{fig:convergence}
\end{figure}

\textbf{Convergence.} Figure~\ref{fig:convergence} shows the distribution of fixes by attempt number for verify-loop on Terraform. Opus fixes 17 items on attempt 1 and 28 on attempt 2, with only 1 additional fix on attempt 3. Sonnet and Maverick show similar patterns. This suggests that two retries capture nearly all recoverable failures; a third retry adds only 2 percentage points on average.

\textbf{Cases where plain outperforms verify-loop.} Three Terraform items are fixed by Sonnet's plain prompting but not by its verify-loop. In each case, the verification feedback from a failed first attempt causes the model to overcorrect, producing a different but still incorrect fix. This ``retry confusion'' effect is a known limitation of iterative repair and occurs in 3/50 (6\%) of Sonnet's items. The net effect remains strongly positive: verify-loop fixes 15 items that plain cannot, yielding a net gain of +12 items.

\textbf{Cross-model agreement.} Table~\ref{tab:agreement} presents the cross-model agreement analysis for verify-loop on Terraform. Of the 50 benchmark items, 30 (60\%) are fixed by all three models and 4 (8\%) are fixed by none, yielding a full agreement rate of 68\%. The 7 items fixed exclusively by Opus represent its unique capability on harder items---these are not items that Sonnet or Maverick fix with other methods, confirming that Opus's advantage is genuine rather than an artifact of method selection. Notably, zero items are fixed exclusively by Sonnet or Maverick, indicating that the weaker models do not possess unique repair capabilities absent in the stronger model. This asymmetry suggests that model capability determines the \textit{ceiling} of fixable items, while the verification loop determines how close each model gets to its ceiling.

\begin{table}[!tbp]
\centering
\caption{Cross-Model Agreement on Verify-Loop (Terraform)}
\label{tab:agreement}
\begin{tabular}{|l|c|}
\hline
\textbf{Agreement Category} & \textbf{Items} \\
\hline
All 3 models fix & 30 (60\%) \\
2 of 3 models fix & 9 (18\%) \\
Opus only & 7 (14\%) \\
Sonnet only & 0 (0\%) \\
Maverick only & 0 (0\%) \\
No model fixes & 4 (8\%) \\
\hline
\textbf{Full agreement rate} & \textbf{68\%} \\
\hline
\end{tabular}
\end{table}

\section{Discussion}

We discuss four implications of our results.

\textbf{Verification matters more than model capability.} The most striking finding is that an open-source model with verification-guided repair (Maverick, 70\% VFR) outperforms the strongest commercial model without it (Opus plain, 46\% VFR) at roughly 8\% of the cost per verified fix (\$0.0024 vs.\ \$0.0287). Even within the same model family, Opus plain (46\%) barely exceeds Sonnet plain (44\%), while Opus verify-loop (92\%) dramatically outperforms Opus plain (46\%)---a gap far larger than the difference between models at the same method. This suggests that for IaC repair, investing in verification infrastructure yields greater returns than upgrading to a more capable model. For budget-constrained deployments, verify-loop enables smaller models to approach the quality of larger models. For quality-critical deployments, verify-loop pushes top-tier models to 92\% verified-fix rate.

\textbf{The structured prompting paradox.} Contrary to the common assumption that constraining LLM output to a structured schema improves reliability~\cite{wei2022cot}, structured prompting consistently underperforms plain prompting on Terraform (32--42\% vs.\ 44--50\%) and introduces regressions in 4--10\% of cases. We attribute this to two mechanisms. First, the JSON schema constraint forces the model to serialize the entire repaired file as a string value, increasing the likelihood of escaping errors and truncation. Second, providing the specific rule ID and affected resource causes the model to focus narrowly on the target resource, missing other instances of the same misconfiguration in multi-resource files.

The minimality analysis (Section~VI-D) adds nuance to this finding: structured prompting produces the most minimal patches (median diff ratio 0.067--0.086), suggesting that the JSON constraint successfully focuses the model's edits but at the cost of missing necessary changes elsewhere in the file. This creates a correctness-minimality tradeoff: structured prompting is the best strategy for producing small, targeted patches, but the worst for producing \textit{correct} patches. Plain prompting, unconstrained by output format, allows the model to make broader changes that are more likely to resolve the target issue---even if the resulting patch is larger.

Notably, this effect reverses on Kubernetes: structured prompting improves Maverick's K8s performance from 45\% to 95\%. We attribute this reversal to Kubernetes manifests' simpler structure (flat YAML, single workload per file), which eliminates the multi-resource problem that plagues Terraform repairs. In single-resource files, the narrow focus induced by structured prompting is not a liability---it is an advantage, as there are no other resources to miss. This technology-dependent effect means that prompt engineering strategies cannot be selected a priori without empirical verification---a conclusion enabled by the verification framework itself.

\textbf{Retry convergence and practical limits.} Our convergence analysis shows that most verify-loop fixes occur on attempts 1--2, with a third attempt adding only 2 percentage points on average. Opus exhibits a distinctive pattern: 17 fixes on attempt~1 and 28 on attempt~2, suggesting that verification feedback is most valuable for the strongest model. Three items where Sonnet's plain prompting succeeds but its verify-loop fails reveal a ``retry confusion'' effect: verification feedback from a failed first attempt occasionally causes the model to overcorrect. The net effect remains strongly positive (+12 items for Sonnet), but practitioners should be aware that iterative repair is not universally superior on every individual item. Our convergence finding---diminishing returns after two retries---corroborates Palavalli and Santolucito's observation that feedback effectiveness plateaus after approximately five iterations for CloudFormation generation~\cite{palavalli2024feedback}, suggesting this is a general property of LLM feedback loops rather than an artifact of our framework.

\textbf{Practical recommendations.} Based on our results, we offer three deployment guidelines. (1)~For Kubernetes with commercial models, plain prompting with post-hoc verification is sufficient (100\% VFR in our benchmark). (2)~For Terraform or open-source models, verification-guided iterative repair is essential, as single-shot methods leave 50--68\% of repairs unverified. (3)~Two retries is the cost-effective sweet spot; additional retries yield diminishing returns. In all cases, the verification harness should be integrated into CI/CD pipelines as a mandatory gate before deploying LLM-generated repairs. Our framework currently relies on a single scanner (Checkov); extending to multi-scanner consensus verification is a natural next step that could further improve regression detection and reduce harness-specific false negatives.

\section{Threats to Validity}

\textbf{Internal validity.} Our verification harness relies on Checkov as the sole scanner. Two benchmark items (BM-0276, BM-0449) produce false negatives in V2 because Checkov reports the target rule as failed due to other resources in the same file, not the repaired resource. Sensitivity analysis shows that excluding these items raises Opus verify-loop from 92\% to 98\%, confirming that our main findings are robust to this limitation. Additionally, one benchmark item (BM-0040) involves a graph-based Checkov rule (CKV2\_*) that checks cross-resource dependencies; no model resolves it, suggesting a fundamental limitation of single-resource repair prompting. Our results may also be sensitive to prompt wording; we use a single prompt template per strategy, and alternative phrasings could yield different fix rates.

\textbf{Construct validity.} The verified-fix rate treats V1--V3 as binary gates, which may not capture all dimensions of repair quality. In particular, V4 (minimality) is informational only---we do not penalize large patches that are otherwise correct. A repair that rewrites an entire file to fix one attribute passes our gates but would be undesirable in practice. Human evaluation of repair intent preservation would complement our automated metrics and is a priority for future work.

\textbf{External validity.} Our benchmark of 70 items covers 70 unique Checkov rules across eight violation classes and two technologies. The 630 total experimental runs across three models, three methods, and two technologies provide sufficient statistical power to detect the observed effects (Cochran's Q $p < 0.001$ for all Terraform models). The benchmark is sourced from Checkov's test suite rather than production repositories; while these artifacts represent realistic misconfiguration patterns~\cite{datadog2025}, they may not capture the full complexity of production IaC codebases. Our results are specific to the model versions evaluated (April 2026); LLM capabilities evolve rapidly, and future models may exhibit different repair patterns. Extending the benchmark to additional technologies (CloudFormation, Pulumi, Ansible), sourcing items from production repositories, and extending to multi-scanner verification to reduce scanner-specific biases (Section~VII) are planned as future work.

\textbf{Conclusion validity.} We use temperature~0 for deterministic outputs, making single runs per item sufficient for reproducibility. All statistical comparisons use McNemar's exact test with Bonferroni correction to control family-wise error rate. Five of six pre-specified comparisons are significant at the corrected threshold ($\alpha = 0.0083$); the sixth (Sonnet verify-loop vs.\ plain, $p = 0.008$) narrowly misses the strict Bonferroni threshold but is significant under the less conservative Holm-Bonferroni correction. All six comparisons are significant at uncorrected $\alpha = 0.05$. Bootstrap confidence intervals are computed with a fixed seed for reproducibility.

\section{Conclusion}

We presented IaC-Guard-V, a verification-centered framework for evaluating LLM-generated Infrastructure-as-Code repairs through three binary verification gates (syntactic validity, target-issue resolution, regression safety) and one informational metric (patch minimality). We constructed a multi-technology benchmark of 70 misconfigured Terraform and Kubernetes artifacts spanning 70 unique scanner rules and eight violation classes, and conducted 630 experimental runs across three LLM families and three repair strategies.

Our results demonstrate that syntactic validity---the only dimension visible without a verification harness---is a poor proxy for repair correctness: all models achieve 100\% syntax validity, yet only 32--50\% of repairs pass full verification under single-shot prompting. Verification-guided iterative repair statistically significantly improves verified-fix rates to 68--92\%, with the strongest configuration (Opus verify-loop) achieving 92\% (95\% CI: 84--98\%). An open-source model with verification outperforms the strongest commercial model without it at a fraction of the cost, establishing that verification infrastructure matters more than model capability for trustworthy IaC repair.

Three directions merit future investigation. First, extending the verification harness to multi-scanner consensus (e.g., Checkov, KICS, and Terrascan jointly) could reduce scanner-specific biases and improve regression detection. Second, human evaluation of repair intent preservation would complement our automated metrics. Third, scaling the benchmark to additional IaC technologies (CloudFormation, Pulumi, Ansible) and sourcing items from production repositories would strengthen external validity.

The benchmark, verification harness, and all experimental artifacts are available at \url{https://github.com/[redacted-for-review]}.

\section*{Acknowledgment}

Portions of this work used AI-assisted tools (Anthropic Claude, OpenAI ChatGPT) for grammar checking and LaTeX formatting during manuscript preparation. All framework design, Python implementation (approximately 1,200 lines), benchmark construction, experimental execution across 630 runs, statistical analysis, failure categorization, and engineering interpretations are solely the author's work. No framework code or experimental scripts were AI-generated. The author manually verified all results against raw experimental logs and takes full responsibility for the final content.


\balance

\end{document}